\documentclass[aps,prl,superscriptaddress,twocolumn,showpacs]{revtex4}

\usepackage{graphicx}
\usepackage{dcolumn}
\usepackage{bm}
\usepackage{subfigure}
\usepackage{color}
\usepackage{hyperref}
\begin{document}

\title{Energy Gap Substructures in Conductance Measurements of MgB$_2$-based
Josephson Junctions: Beyond the 2-Gap Model }

\author{Steve Carabello}
\email{sac69@drexel.edu}
\affiliation{Drexel University, Philadelphia, PA 19104 USA}
\affiliation{Penn State Harrisburg, Middletown, PA 17057, USA}

\author{Joseph G. Lambert}
\author{Jerome Mlack}
\affiliation{Drexel University, Philadelphia, PA 19104 USA}

\author{Wenqing Dai}
\author{Qi Li}
\affiliation{The Pennsylvania State University, University Park, PA 16802, USA}
\author{Ke Chen}
\author{Daniel Cunnane}
\affiliation{Temple University, Philadelphia, PA 19122, USA}

\author{C.G. Zhuang}
\affiliation{The Pennsylvania State University, University Park, PA 16802, USA}
\affiliation{Temple University, Philadelphia, PA 19122, USA}

\author{X. X. Xi}
\affiliation{Temple University, Philadelphia, PA 19122, USA}

\author{Roberto C. Ramos}
\affiliation{Indiana Wesleyan University, Marion, IN 46953 USA}

\date{\today}

\begin{abstract}
Several theoretical analyses of the two superconducting energy gaps of
magnesium diboride, $\Delta_\pi$ and $\Delta_\sigma$,
predict substructures within each energy gap, rather than two pure numbers.
Recent experiments have revealed similar structures. We report tunneling
conductance data providing additional experimental evidence for these features.
The absence of these features in $c$-axis tunneling, and a sharp peak in the
subgap (associated with the counterelectrode material), support the conclusion
that these features are intrinsic to MgB$_2$. By demonstrating the inadequacy
of a simple two-gap model in fitting the data, we illustrate that some
distinctions between theoretical models of energy gap substructures are
experimentally accessible.
\end{abstract}

\maketitle

\section{Introduction}

Magnesium diboride (MgB$_2$) has a number of properties making it a
particularly interesting object of study. Among them are its two well-separated
energy gaps. Although there had long been both experimental
\cite{PhysRevB.2.4511,PhysRevB.5.1171} and theoretical
\cite{PhysRevLett.3.552,Rogovin1976175,Schopohl1977371} suggestions of two-gap
superconductivity, MgB$_2$ was the first material to put the matter beyond
dispute\cite{0034-4885-71-11-116501}. Multi-gap superconductivity has recently
attracted increased interest, with its demonstration in a variety of materials,
including pnictides \cite{0953-2048-24-4-043001}.

As theories were developed for understanding superconductivity in MgB$_2$, it
was recognized that its superconducting energy gap must be both anisotropic and
multi-valued: the higher energy gap is associated with the strong $\sigma$
bonds in the Boron planes, and the lower energy gap is associated with the
weaker $\pi$ bonds.

\begin{figure}[htbp]
		\includegraphics[scale=1]{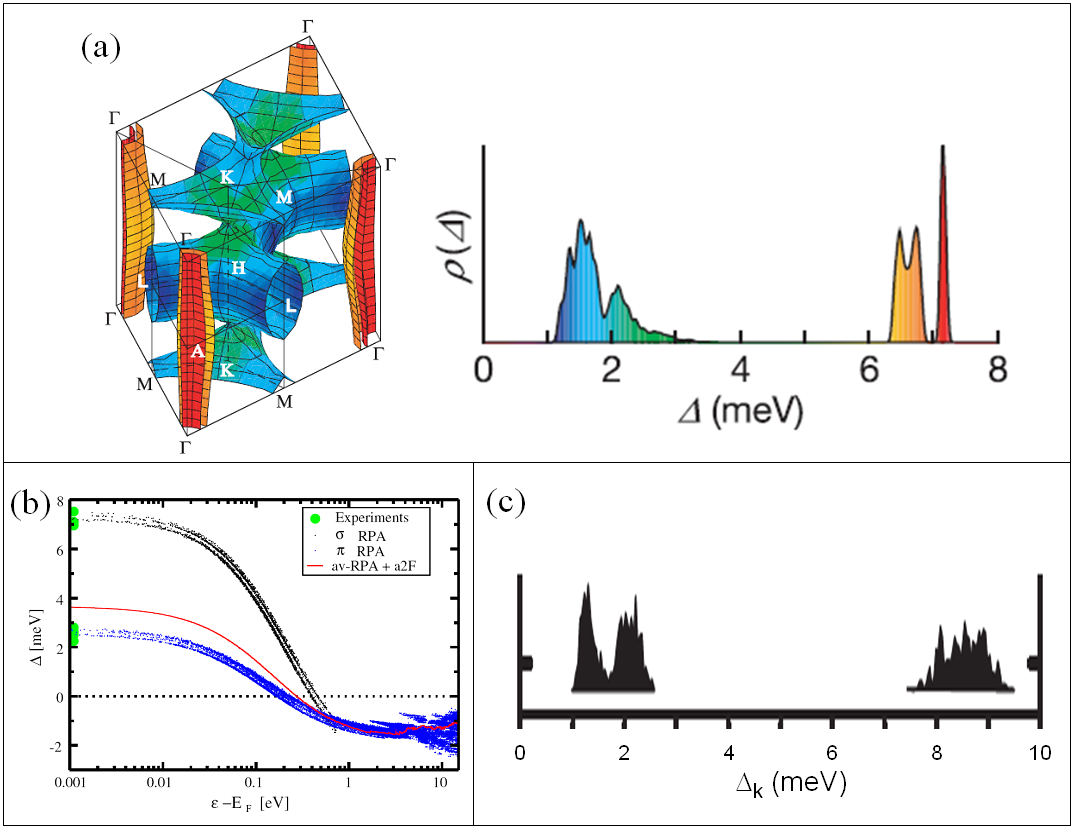}
\caption{ \label{Figure1}(color online) Theoretical models for the energy gap
distribution of MgB$_2$. (a) Fermi surface and corresponding gap structure, at
$T$=0 (Adapted by permission from Macmillan Publishers Ltd: Nature 418, 758 $\copyright$ 2002 \cite{choi-2002-418}). (b) The superconducting energy
gap as a function of energy distance from the Fermi energy at $T=0$ (reproduced
from \cite{PhysRevLett.94.037004}). (c) Calculated anisotropic superconducting
energy gaps, at low temperature (extracted from \cite{PhysRevB.87.024505}). }

\end{figure}

Several theoretical analyses revealed sub-features in each energy gap,
reflecting the electron-phonon interactions in MgB$_2$. Figure \ref{Figure1}(a)
shows the Fermi surface of MgB$_2$ together with the corresponding local
density of states at each gap energy \cite{choi-2002-418}. This model was among
the earliest to show the distribution in gap energies explicitly. In Figure
\ref{Figure1}(b), the computed energy gap as a function of energy from the
Fermi energy is shown \cite{PhysRevLett.94.037004}. Near the Fermi level
($\epsilon$-E$_F$=0), it also reveals a distribution rather than a single
energy, for each gap. A more recent model (Figure
\ref{Figure1}(c))\cite{PhysRevB.87.024505} also exhibits a distribution in the
superconducting gaps of MgB$_2$. Although each model determines the gap
distribution from first principles, differing assumptions and parameter values
are applied. These lead to differences in the features of the gap
distributions, and the energies at which they appear.

A variety of measurements have provided evidence for two-gap superconductivity
in MgB$_2$ \cite{0034-4885-71-11-116501,PhysRevLett.87.157002}, with
conductance curves from tunneling and point-contact spectroscopy proving to be
particularly useful techniques \cite{PhysRevLett.89.187002,
0953-2048-23-4-043001,PhysRevLett.89.247004,Ponomarev200485,
Schmidt2003221}. As expected, tunneling in the $a$-$b$ plane shows strong 
contributions from both the $\pi$ and $\sigma$ gaps, while tunneling along the 
$c$-axis primarily exhibits the $\pi$ gap.

Features consistent with substructure within each energy gap have been found in
some experimental data \cite{ncomms1626, ISI:000315667500054,
chen:012502,10.1134_1.1780557, 5643942}. These fine features had been thought
to be unobservable in physically-realistic systems \cite{PhysRevB.69.056501}.
Their observation opens a new avenue for exploring superconductivity in
MgB$_2$.

We have conducted differential conductance measurements on MgB$_2$/I/Pb and
MgB$_2$/I/Sn junctions with a variety of film geometries. Above $T_c$ of the Pb
or Sn electrode, our results are consistent with a simple two-gap model.
However, below $T_c$, our results distinguish simplified two-gap and four-gap
models. Therefore, we demonstrate the need to go beyond the two-gap model.

\section{Simple 2-Gap and 4-Gap Theory}

We begin our analysis with the current-voltage characteristics of a generalized
junction between two materials,

\begin{equation}
\label{equation1}
I(V)=G_n \int^\infty_{-\infty} N_1(E)N_2(E+eV)[f(E)-f(E+eV)]\,dE
\end{equation}
\noindent where $G_n$ is the normal-state conductance of the junction (assumed
constant), $N(E)$ is the density of states for each electrode, and $f(E)$ is
the Fermi distribution.

Using the BCS density of states for each superconducting electrode,
\begin{equation}
\label{equation2}
N(E)=\Re\left\{\sqrt{ \frac{E^2}{E^2-\Delta^2}}\right\}
\end{equation}
\noindent one is able to reproduce the current-voltage characteristics of a
basic tunnel junction (neglecting the Josephson supercurrent). Since we measure
junctions made from two different electrode materials, each density of states
will use a different energy gap $\Delta_1$ and $\Delta_2$, where $\Delta_1$
($\Delta_2$) represents the lower (higher) of the two gaps.

In the simplified model we have used (matching that described in
\cite{0953-2048-23-4-043001}), two additional effects are considered:

\textbf{Broadening Factor $\mathbf{\Gamma}$}: Dynes \textit{et
al.}\cite{PhysRevLett.41.1509} found a broadening in conductance peaks that
could not be attributed to temperature. Instead, the quasiparticle lifetime
provides a broadening that can be accounted for by replacing all instances of
$E$ with $E+i\Gamma$ in the BCS density of states.

$\Gamma$ has also been used to simulate the effect of a convolution of the
theoretical conductance with a distribution of gap
values\cite{0953-2048-23-4-043001}. In this paper, we model the gap
distribution as distinct gap energies, with a broadening that
phenomenologically matches the experimental gap distribution from our
experiments.

Including a constant $\Gamma$ reveals a feature in the subgap region of the
$I-V$ curve, which we have observed and used in our analysis. With a constant
$\Gamma$ included, the modified BCS density of states is nonzero, even at
$E=0$, whether or not the transparency of the junction is zero. This allows the
formation of peaks at $\Delta_1$ and $\Delta_2$ in the subgap region, down to
$T=0$, in the absence of Andreev reflections (which require a finite
transparency). Because of their strong sensitivity to thermal broadening, the
peaks virtually disappear above $3K$ in theoretical calculations using values
similar to those of our junctions. But, the peak at $\Delta_1$ becomes quite
sharp as $T\rightarrow 0$.

\textbf{Weighting for multiple gaps}: An additional refinement must be made
when considering a multi-gap superconductor. For a two-gap model, a single
weighting factor is used:
\begin{equation}
\label{equation3}
N(E)=w_1N_1(E)+(1-w_1)N_2(E)
\end{equation}
\noindent If additional peaks are observed, they can be modeled as additional
gaps, each with its own gap energy $\Delta$, its own broadening factor
$\Gamma$, and its own weight $w$, as long as the sum of the weights equals 1.

For MgB$_2$, two gaps are usually assumed. The weighting factors depend on
junction geometry, with $w_{\sigma}=1-w_{\pi}$ ranging from less than 1\% for
pure $c$-axis tunneling, to $\sim33\%$ for pure $a$-$b$ plane
tunneling\cite{0953-2048-23-4-043001}.

As indicated in Figure \ref{Figure1}, two smoothly-broadened gaps (one each for
$\pi$ and $\sigma$) may not be sufficient to represent the density of states of
MgB$_2$. As a result, \emph{within the $\pi$ (or $\sigma$) gap, substructures
are necessary, each with its own weighting factor}. For simplicity, we have
chosen to model each structure as its own gap, with its own broadening.

From the Fermi surface of MgB$_2$, it is evident that the gap value within the
$\pi$ gap will be dependent on the tunneling direction. Therefore, the weights
for each substructure should differ for different samples.

\begin{figure}[htbp]
		\includegraphics[scale=1]{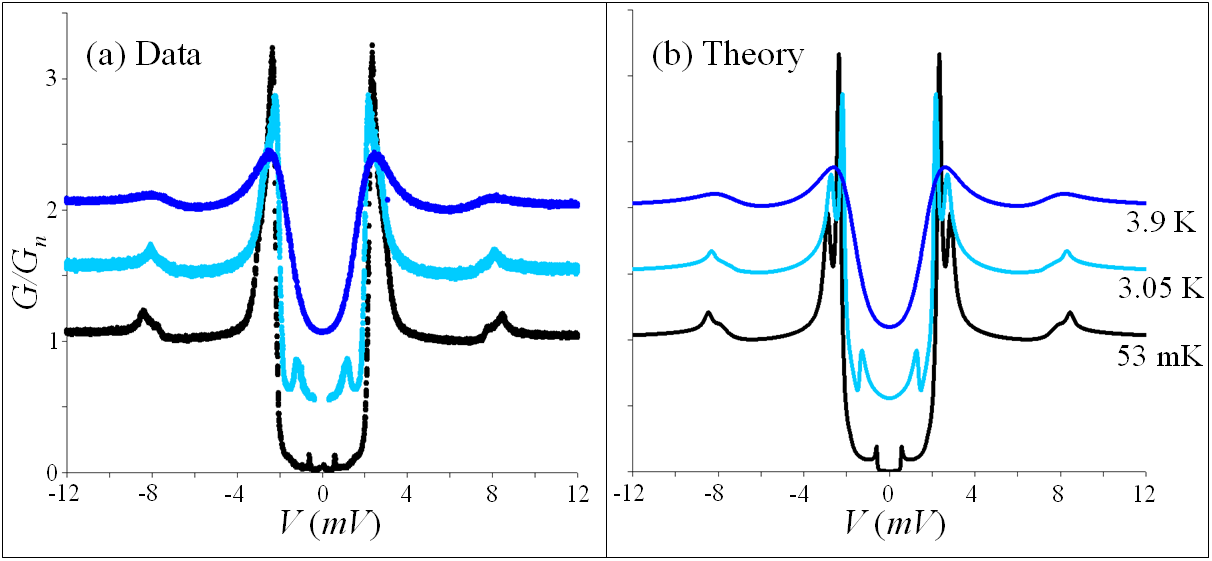}
\caption{ \label{Figure2}(color online) (a) Experimental and (b) theoretical
normalized conductance vs. voltage curves at 53mK, 3.1K, and 3.9K of an
MgB$_2$/I/Sn junction. Curves have been offset for clarity. The ability to
resolve features improves dramatically as the junction transitions from
$T>T_{c\,Sn}$ (forming an $NS$ junction) to $T<T_{c\,Sn}\sim$3.7K (forming an
$SIS'$ junction). Above $T_{c\,Sn}$, 2-gap and 4-gap models fit the data
equally well.}

\end{figure}

We have chosen to compare and contrast our data with a simple 2-gap model (one
gap for $\pi$ and one for $\sigma$) vs. a model consisting of four gaps (two
each for $\pi$ and $\sigma$). We have observed that, below $T_c$ of the Pb or
Sn counterelectrode, a 4-gap model is superior. Above that temperature, 2-gap
and 4-gap models cannot be distinguished (see Figure \ref{Figure2}).

Hereafter, we will refer to these gaps as $\pi_1$, $\pi_2$, $\sigma_1$, and
$\sigma_2$, with subscripts 1 (2) referring to the peak at the lower- (higher-)
energy sub-peaks of the $\pi$ and $\sigma$ gaps. We assume a single
sample-dependent gap value $\Delta$, broadening parameter $\Gamma$, and weight
$w$ for each gap. We emphasize that we are not suggesting a physical source for
these gap values. Instead, they serve as a convenient model for the gap
distribution.

\textbf{Neglected Effects}: We use a simplified model because we do not assume
that any one of the existing theoretical models of the gap distribution is
correct. Several additional physical effects have been neglected in our
analysis (some of which are described below), either because they have a minor
effect on our data, or to limit the number of free parameters to a manageable
level.

All of the junctions measured in this study were good tunnel junctions with
strong barriers. As a result, there should be little contribution from Andreev
reflections, computed using the OBTK
model\cite{PhysRevB.27.6739,PhysRevB.38.8707,PhysRevB.54.6557,
PhysRevB.56.11232}.

In superconductors with strong electron-phonon coupling, $\Delta$ must be
treated as a complex function of energy, rather than a constant value. However,
theoretical \cite{PhysRevB.87.024505,PhysRevB.68.132503,PhysRevB.69.100501} and
experimental \cite{PhysRevLett.94.227005} studies find that their contributions
for MgB$_2$ are small for $|E| \lesssim 30$ meV. We neglect this effect, since
our study seeks features within the $\pi$ and $\sigma$ gaps ($|E| \leq 15$
meV).

In a two-band superconductor, quasiparticles may scatter from one band to the
other, leading to coupled energy-dependent gap functions
$\Delta(E)$\cite{Schopohl1977371,PhysRevB.82.014531,PhysRevB.81.104522}. In a
two-gap model, this introduces two additional parameters. For a 4-gap model,
the number of free parameters would grow unwieldy, without assisting us in our
goal of demonstrating the inadequacy of a two-gap model. As a result, we
neglect this effect as well.

\section{Experimental Design}

Tunneling conductance curves are well-suited for determining the energy gaps
and densities of states of superconducting materials. The differential
conductance of an $N$-$S$ junction at $T$=0 is proportional to the density of
states of the superconductor; at finite temperature, it is smeared by
$\sim\pm2kT$\cite{nla.cat-vn1834745}. In $SIS'$ junctions, the ``very sharply
peaked densities of states at the gap edges of both materials helps to
counteract the effects of thermal smearing.'' \cite{nla.cat-vn1834745} (See
Figure \ref{Figure2}.) By using SIS' junctions at mK temperatures,
incorporating extremely high-quality MgB$_2$ films, we maximize our ability to
resolve features within the energy gaps.

\begin{figure}[htbp]
		\includegraphics[scale=1]{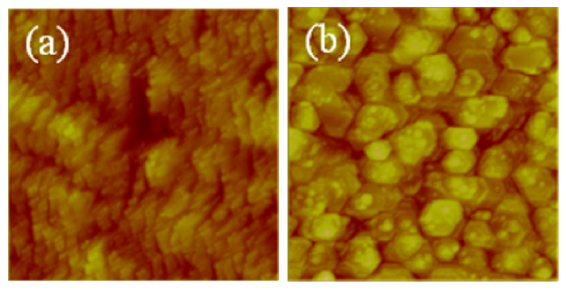}
\caption{ \label{Figure3}(color online) AFM images of representative MgB$_2$
films on SiC (Reprinted with permission from J. Appl. Phys. 113, 083902. Copyright 2013, AIP Publishing LLC. \cite{ISI:000315667500054}). Each
image represents a 2$\mu$m $\times$ 2$\mu$m area. (a) A ``terraced'' film on
$8^{\circ}$ SiC, similar to that used for the MgB$_2$/I/Sn junction discussed
below. (b) A ``columnar'' film on rough $c$-axis SiC, similar to that used for
the MgB$_2$/I/Pb junction discussed below.}

\end{figure}

For this study, we used MgB$_2$/I/Pb and MgB$_2$/I/Sn tunnel junctions
incorporating high purity MgB$_2$ thin films grown by hybrid physical-chemical
vapor deposition (HPCVD) on single-crystal SiC substrates\cite{chen:012502}. As
found by Dai $et.\,al.$ \cite{ISI:000315667500054}, on smooth $0^{\circ}$ SiC,
a largely-planar MgB$_2$ film forms, exposing primarily the $c$-axis for
tunneling. On SiC whose polished surface is tilted $8^{\circ}$ from the
$c$-axis, the MgB$_2$ film takes on a ``terraced'' shape, exposing the $a$-$b$
plane. On rough $0^{\circ}$ SiC, the growth of the MgB$_2$ film forms columnar
structures, exposing even more of the $a$-$b$ plane, while still maintaining
clean high-quality films. (See Figure \ref{Figure3}.)

The insulating barrier is formed by a native oxide, which forms upon exposure
of the film to air, and creates a good tunnel barrier under proper conditions.
A Pb or Sn counterelectrode $\sim$0.3mm wide is then thermally evaporated on a
$\sim$0.3mm exposed strip of the film.

These junctions are cooled in a Helium dilution refrigerator with a base
temperature $\sim$20mK.

The current bias for our junction was provided by sweeping the voltage from an
Agilent 3220A function generator, through a bias resistor, prior to reaching
the junction. By ensuring $R_{bias} \gg R_{junction}$, this combination behaves
effectively as a current source. The resulting voltage across the junction was
amplified, then recorded. The current bias was swept at between 10 mHz and 1
Hz, while measurements were acquired from 10 kHz to 48 kHz. Such oversampling
allows numerical differentiation to produce high-resolution results, by
averaging adjacent data points. The number of points in each average was
proportional to the time spent near any given voltage. All results were robust
under a variety of averaging methods, and a number of results were verified
using an SR830 lock-in amplifier, demonstrating that the features are not an
artifact of the averaging process.

Electrical isolation of the cryostat was provided by Stanford Research 560
amplifiers operating in differential mode. Additionally, the conductive path
through the vacuum pumping lines was broken using plastic clamps and centering
rings. High-frequency signals were filtered via thermally-grounded Thermocoax
cables, followed by LC and copper powder filters mounted to the cold finger.
Magnetic fields were excluded via cryoperm shielding, and below $\sim$1K, the
Aluminum sample box expels magnetic fields. Vibration damping pillars supported
the cryostat. For most measurements\cite{Note_NI9215}, a National Instruments
(NI) 9239 DAQ was used to acquire data and store it to the computer. Its inputs
are well-isolated, with minimal crosstalk.

Voltage amplification for the ``columnar'' junction was provided by an SR 560
amplifier operating in differential mode. For the ``terraced'' and $c$-axis
junctions, a home-built battery-powered amplifier using 4 JFETs in parallel was
used.\cite{XuPhd2004,BerkeleyPhd2003} The high input impedance of the JFETs
ensures that very little current flows along the voltage measurement lines, and
severely limits the ability of noise signals to return to the junction.

\section{Key Results}
\subsection{Obtaining $\Delta_{Pb}$ and $\Delta_{Sn}$ from Subgap Features}

\begin{figure}[htbp]
		\includegraphics[scale=1]{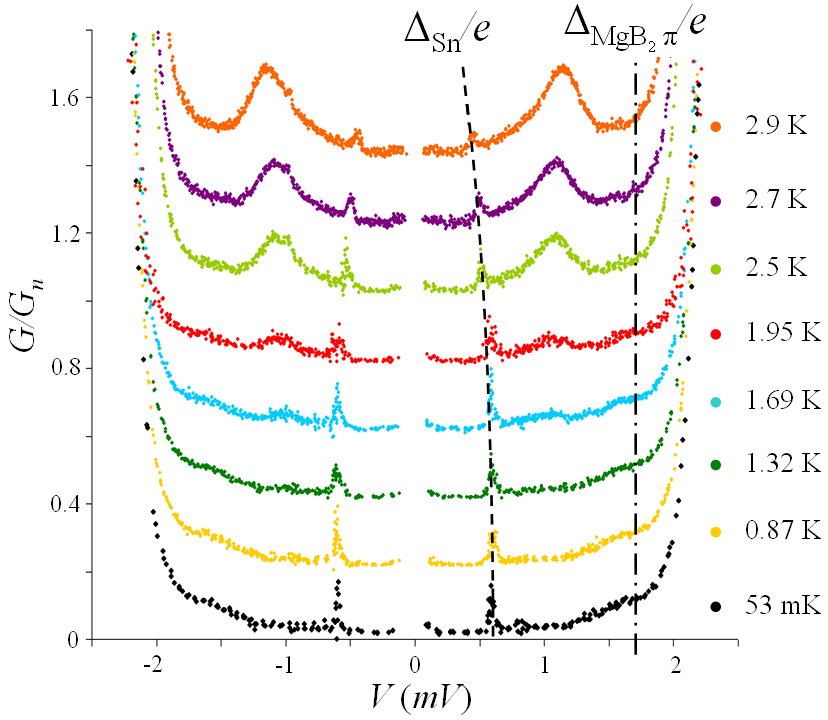}
\caption{ \label{Figure4}(color online) Conductance data in the subgap region,
for a ``terraced'' MgB$_2$/I/Sn junction with $R_n = 15\Omega$ and
$R_{sg}\gtrsim 600\Omega$, from 53 mK to 2.9K. Curves have been offset for
clarity. The sharp peak at $\Delta_{Sn}/e$ is used to establish $\Delta_{Sn}$
in our analysis.}

\end{figure}

\begin{figure}[htbp]
		\includegraphics[scale=1]{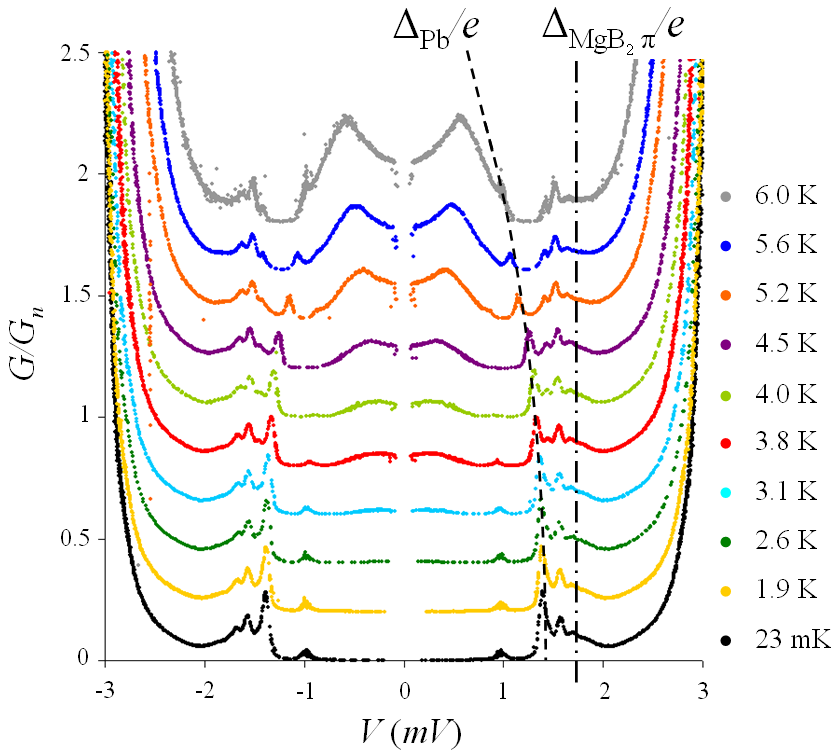}
\caption{ \label{Figure5}(color online) Conductance data in the subgap region,
for a $c$-axis MgB$_2$/I/Pb junction with $R_n = 104\Omega$ and $R_{sg}\gtrsim
13k\Omega$, from 23mK to 6K. The most prominent low-temperature peak appears at
$\Delta_{Pb}/e$. The sharpness of this peak, and the small conductance at
voltages below this peak, are indications of a small $\Gamma_{Pb}$. A broader
peak at $\Delta_{MgB_{2\,\pi}}/e$ is also evident, exhibiting far less
temperature dependence, which is expected for MgB$_2$ ($T_c = 39$K). Additional
peaks are also evident, which will be discussed in a future article.}

\end{figure}

As discussed above, when a nonzero $\Gamma$ is used, conductance peaks are
expected to appear at the gap voltages $\Delta/e$ of each superconductor. These
are extremely useful, for three reasons. First, they allow us to determine the
gap energy of the Pb and Sn counterelectrode to high precision, particularly as
$T\rightarrow 0$. Second, because these subgap peaks are sharp and narrow, we
are confident that the features we observe in the
$(\Delta_{MgB_2}+\Delta_{Sn/Pb})/e$ peaks are due to MgB$_2$ rather than the
counterelectrode material. Finally, they establish that $\Gamma$ for the
counterelectrode material is small, which reduces the parameter space being
explored by our models.

Many of the observed features are reasonably consistent with the simple 2-gap
(one for $\pi$ and one for $\sigma$) and 4-gap (two each for $\pi$ and
$\sigma$) models described above, as can be seen in the subgap portions of
Figures \ref{Figure7}, \ref{Figure8}, and \ref{Figure9}. However, a more
sophisticated model is required to completely reproduce all of the features. A
full discussion is beyond the scope of this paper, and will be addressed in a
future article.

We also note that the peak at $\Delta_{MgB_{2}\,\pi}/e$ is quite broad, which
is appropriate given the distribution in gap values expected in $MgB_{2}$.

At higher $T$, the peak at $(\Delta_1 - \Delta_2)/e$ appears. This peak is due
to quasiparticles thermally excited across the energy
gap\cite{nla.cat-vn1834745}. The well-defined peaks at both $(\Delta_1 +
\Delta_2)/e$ and $(\Delta_1 - \Delta_2)/e$ are widely used to find unique
numerical values for both gaps. In this case, however, the $(\Delta_1 -
\Delta_2)/e$ peak takes on a rounded appearance due to the distribution in the
$\pi$ gap energies of MgB$_2$, in addition to thermal broadening. Therefore, we
use the subgap peaks at $\Delta_{Sn/Pb}/e$ together with the peaks at
$(\Delta_1 + \Delta_2)/e$ to acquire unique values for each gap.

\subsection{Calculating Gap Weights}

When properly normalized, both theoretical and experimental conductance curves
must approach 1 as $V$ approaches infinity. For low-transparency junctions with
moderate broadening, it typically becomes quite close to 1 above roughly twice
the gap voltage $V_g = (\Delta_1+\Delta_2)/e$. If a two-gap model is used (one
for $\pi$ and one for $\sigma$), then the curve should approach $w_\pi$ for $eV
\gtrsim 2(\Delta_{Pb}+\Delta_\pi)$, and only reach 1 above
$(\Delta_{Pb}+\Delta_\sigma)$. If a four-gap model is used (two for $\pi$ and
two for $\sigma$), then the curve should approach $(w_{\pi_1} + w_{\pi_2})$ for
$eV \gtrsim 2(\Delta_{Pb}+\Delta_{\pi\,1})$. Therefore, the weight of the $\pi$
gap (or the sum of the weights of the $\pi$ sub-gaps) can be established to
good precision, from data; the weight of the $\sigma$ gap (or the sum of the
weights of the $\sigma$ sub-gaps) will then be $1-w_\pi$.

\begin{figure}[htbp]
		\includegraphics[scale=1]{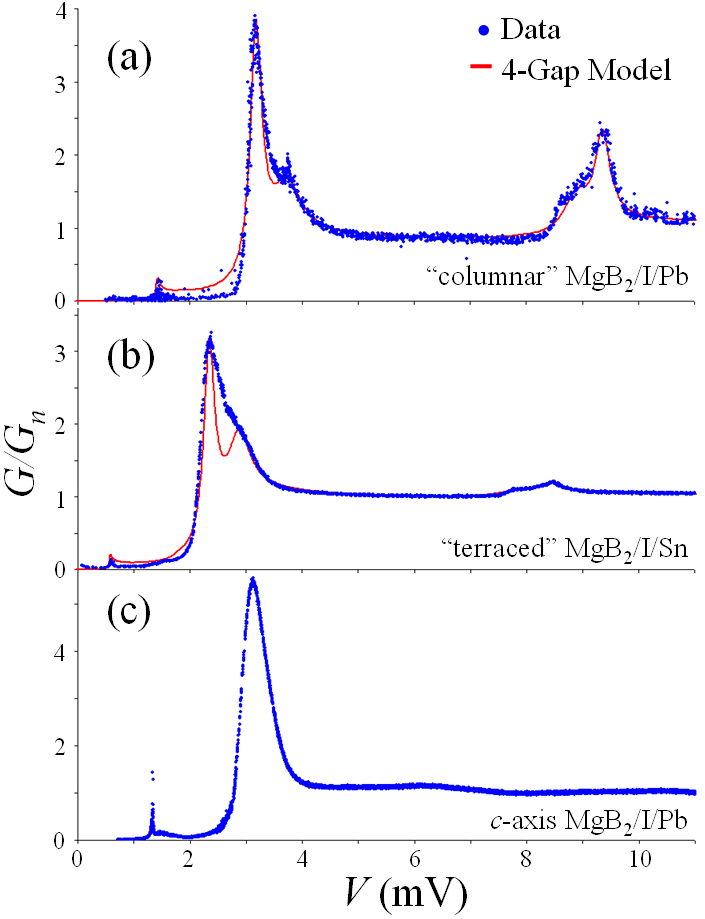}
\caption{ \label{Figure6}(color online) Normalized conductance data and models,
for two different electrode materials and three film geometries. (a)
MgB$_2$/I/Pb results with a ``columnar'' MgB$_2$ film. $w_\sigma\sim$20\%,
indicating significant tunneling along the $a$-$b$ plane. (b) MgB$_2$/I/Sn
results, with $w_\sigma\sim$6\%. (c) MgB$_2$/I/Pb results with a
planar``$c$-axis'' MgB$_2$ film. The peaks are shifted in voltage due to the
difference between the energy gaps of lead ($\Delta_{Pb}\simeq$1.4meV) and tin
($\Delta_{Sn}\simeq$0.57meV). The ``shelf'' feature at $\sim$7mV in the
MgB$_2$/I/Pb data, and one at $\sim$11mV (not shown), are associated with peaks
in the Eliashberg spectral function $\alpha^{2}F(E)$ of the strongly-coupled
superconductor Pb.\cite{wolf2011principles}}

\end{figure}

Applying this method to our ``columnar'' MgB$_2$/I/Pb junction, we find
$w_\sigma \simeq 20\%$. This value is remarkably high, considering that the
MgB$_2$ film was deposited on $0^\circ$ SiC, and the theoretical maximum for
pure $a$-$b$ plane tunneling is $\sim33\%$. The ``terraced'' MgB$_2$/I/Pb
junction we studied has $w_\sigma \sim6\%$. And, as expected, for tunneling to
a pure c-axis MgB$_2$ film, the $\sigma$ peak was indistinguishable
($w_{\sigma}< 1\%$).

\subsection{$\pi$ Gap Substructure}

A majority of Cooper pairs tunneling into an MgB$_2$ surface are expected to
tunnel to the $\pi$ gap. However, the precise details will be sample-dependent.
Here, we consider the three significantly different film geometries described
above.

``$\bf{Columnar}$'': On a rough $0^{\circ}$ SiC substrate, an MgB$_2$ film was
grown, which formed columnar structures (Figure \ref{Figure3} (b)). Pb was
thermally evaporated as the counterelectrode. Since the MgB$_2$ crystallites
were far smaller than the area of Pb in contact with the film, tunneling could
occur along the $c$-axis, along the $a$-$b$ plane, and anywhere in between.
This allows the entire Fermi surface to be explored simultaneously.

\begin{figure}[htbp]
		\includegraphics[scale=1]{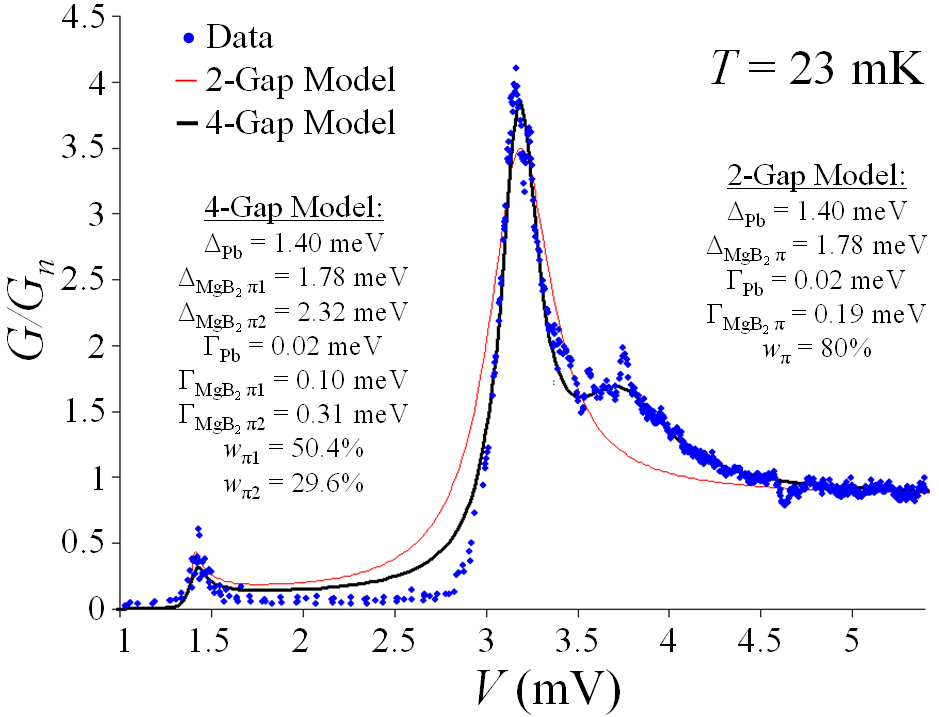}
\caption{ \label{Figure7}(color online) Normalized conductance, showing the
$\pi$ gap for the MgB$_2$/I/Pb ``columnar'' junction. This data is well-modeled
using two $\pi$ gaps at 1.78 and 2.32 meV (with additional gaps for sigma),
while a single $\pi$ peak is unable to capture significant portions of the
data.}

\end{figure}

The gap distributions shown in Figure \ref{Figure1}(a) and
(c)\cite{choi-2002-418,PhysRevB.87.024505} suggest that the $\pi$ gap has a
double-peaked structure. Low-temperature data for this ``columnar'' sample,
shown in Figure \ref{Figure7}, as predicted, displays a double peak at the gap
voltage ($\Delta_{MgB_{2\,\pi}}+\Delta_{Pb})/e$.

Clearly, a model possessing a single $\pi$ gap cannot reproduce this structure.
However, a simple model reflecting features for both $\pi_1$ and $\pi_2$ (in
addition to multiple $\sigma$ gaps, described below) produces remarkable
agreement.

The peak in the subgap region establishes $\Delta_{Pb}$, while the level of the
normalized conductance between $\sim$5mV and $\sim$8mV establishes $w_\pi$ in a
2-gap model, or $\left( w_{\pi\,1} + w_{\pi\,2} \right)$ in a 4-gap model.
Using the taller peak at $\left( \Delta_{\pi\,1} + \Delta_{Pb} \right)/e$,
$\Delta_{MgB_{2}\pi1}$ was determined to a high precision: 1.78meV with an
uncertainty of $\pm$0.02meV. (All uncertainties are estimated by finding a
range of parameters that produce reasonable fits, similar to the method
outlined in \cite{0953-2048-23-4-043001}.)

The remaining parameters in the 4-gap model are less certain. Given the
asymmetry of the $\pi_2$ shoulder, reasonably good 4-gap fits yield
$\Delta_{MgB_{2}\pi2}$ of 2.32meV with an uncertainty of $\sim\pm0.1$meV. The
weights $w$ and the broadening parameters $\Gamma$ are more uncertain, because
according to the model, any peak may have its height decreased by increasing
$\Gamma$ or by decreasing $w$, and vice versa. However, as seen in Figure
\ref{Figure1}, there are not two simple sharp peaks for the $\pi$ gap; and the
broadening parameter is here being used exclusively as a means of approximating
a distribution in the gap energies. So, although $w$ and $\Gamma$ necessary
parameters in the fit, they do not affect our goal in establishing the need for
more than a single broadened $\pi$ gap.

Therefore, we have demonstrated that a single gap energy for the $\pi$ gap is
quite far from the real behavior of MgB$_2$, while a $\pi$ gap possessing two
distinct sub-bands is a reasonable approximation.

``$\bf{Terraced}$'': The MgB$_2$/I/Sn ``terraced'' junction used an MgB$_2$
film formed of parallel tilted layers, each exposing a portion of the $a$-$b$
plane as well as the $c$-axis.

\begin{figure}[htbp]
		\includegraphics[scale=1]{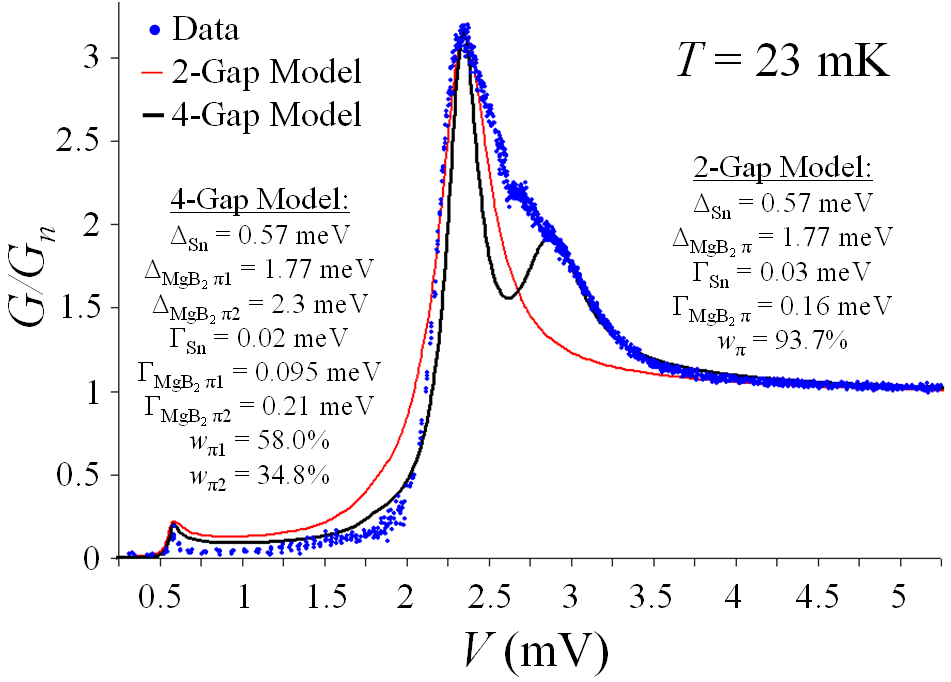}
\caption{ \label{Figure8}(color online) $\pi$ gap for the MgB$_2$/I/Sn
``terraced'' junction. The prominent shoulder at $\sim$3mV indicates that a
single gap energy is not appropriate for the $\pi$ gap. A better fit is given
by a four-gap model with $\pi$ gap values of $\pi_1$ = 1.77meV and $\pi_2$ =
2.3 meV.}

\end{figure}

Once again, a single $\pi$ gap model fits this data very poorly. As shown in
Figure \ref{Figure8}, our simple model including $\pi_1$ and $\pi_2$ is less
successful than for the ``columnar'' sample, though it still does capture some
of the main features. Here, fitting to the tallest peak yields a $\pi_1$ gap
voltage of 1.77$\pm$0.04 meV. However, the broad shoulder can be fit by a wide
range of $\pi_2$ gap energies, with an uncertainty of $\sim\pm$0.2meV.

Within our model, it was not possible to faithfully fit the data, as shown in
Figure \ref{Figure8}. This suggests that a more sophisticated model, capable of
accounting for additional sample-dependent effects, is required. However, we
have demonstrated that a single $\pi$ gap, no matter how much it is broadened,
is inconsistent with the data.

$\bf{\emph{c}-axis}$: In pure $c$-axis tunneling, almost all of the tunneling
is to the $\pi$ gap, with minimal contribution from the $\sigma$ gap. Moreover,
the distribution within the $\pi$ gap should be more limited than in cases
where the $a$-$b$ plane is exposed, since less of the Fermi surface is being
explored.

\begin{figure}[htbp]
		\includegraphics[scale=1]{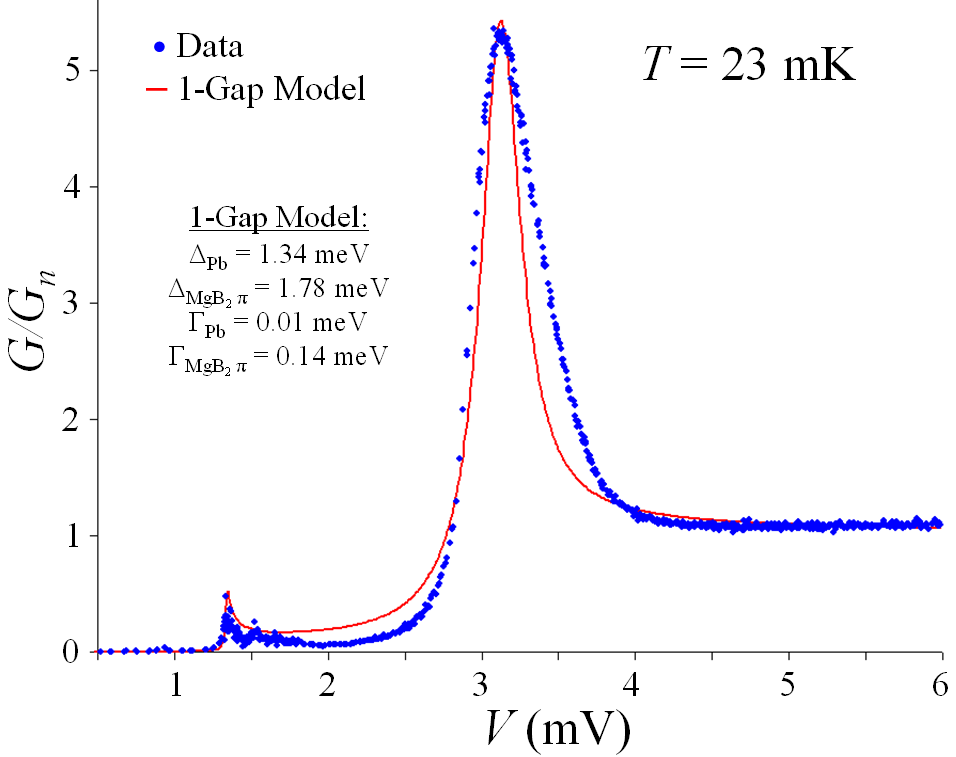}
\caption{ \label{Figure9}(color online) $\pi$ gap for a MgB$_2$/I/Pb c-axis
junction. With tunneling purely along the c-axis of MgB$_2$, far less of the
Fermi surface is being explored. As a result, the gap distribution is far less
apparent. Even so, a single $\pi$ gap (broadened via $\Gamma$) is unable to
match the data.}

\end{figure}

Low-temperature data on a $c$-axis MgB$_2$/I/Pb junction is shown in Figure
\ref{Figure9}. It consists mainly of a single peak, centered at the
lower-energy $\pi_1$ sub-gap. (That is, the peak appears at a voltage
$(\Delta_{MgB_{2\,\pi_1}}+\Delta_{Pb})/e$.) However, using a single-gap theory
with variable $\Gamma$, it was not possible to find a combination of $\Delta$
and $\Gamma$ which make the peak broad enough to match the data.

From this data alone, it is not clear whether a distribution in gap energies is
required, or if a different broadening (which cannot be modeled using $\Gamma$)
is sufficient.

However, we gain increased confidence in the significance of features observed
in the other film geometries because, as expected, this junction shows a more
limited distribution. Each film was fabricated using similar methods, and all
were deposited on SiC. So, it is reasonable to expect that any artificial
contributions due to variations in film strain or other effects should have
appeared in these $c$-axis samples as well. That they did not supports our
suggestion that they are inherent properties of MgB$_2$.

\subsection{$\sigma$ Gap Substructure}

The high $T_c$ of MgB$_2$ is due to the Cooper pairs forming in the $\sigma$
band. Therefore, understanding the $\sigma$ gap is of key importance in
theoretical models. As shown in Figure \ref{Figure1}, there are appreciable
differences between different models, so high-resolution gap measurements may
be of value.

As seen in Figure \ref{Figure6}(c), tunneling to a $c$-axis MgB$_2$ surface
naturally shows no features in the $\sigma$ gap. However, the other contact
geometries do produce useful information.

The ``columnar'' MgB$_2$/I/Pb junction data exhibits features that are clearly
incompatible with a simple 2-gap model. If there is only a single $\sigma$ gap,
then the resulting curve must take on the shape of a broadened BCS density of
states: a steeper low-energy edge, and a gradual decay toward its limiting
value of 1 at higher energies. Our data reveals exactly the opposite: a
relatively sharp peak at high energies, together with a prominent lower-energy
shoulder. These features are reasonably well-modeled by two separate $\sigma$
gaps (which, together with the two $\pi$ gaps form a 4-gap model). However,
there are some features that cannot be fit by two sub-gaps within $\sigma$.
Indeed, the theoretical models shown Figure \ref{Figure1} suggest that two
peaks are insufficient to accurately portray the $\sigma$ gap.

\begin{figure}[htbp]
		\includegraphics[scale=1]{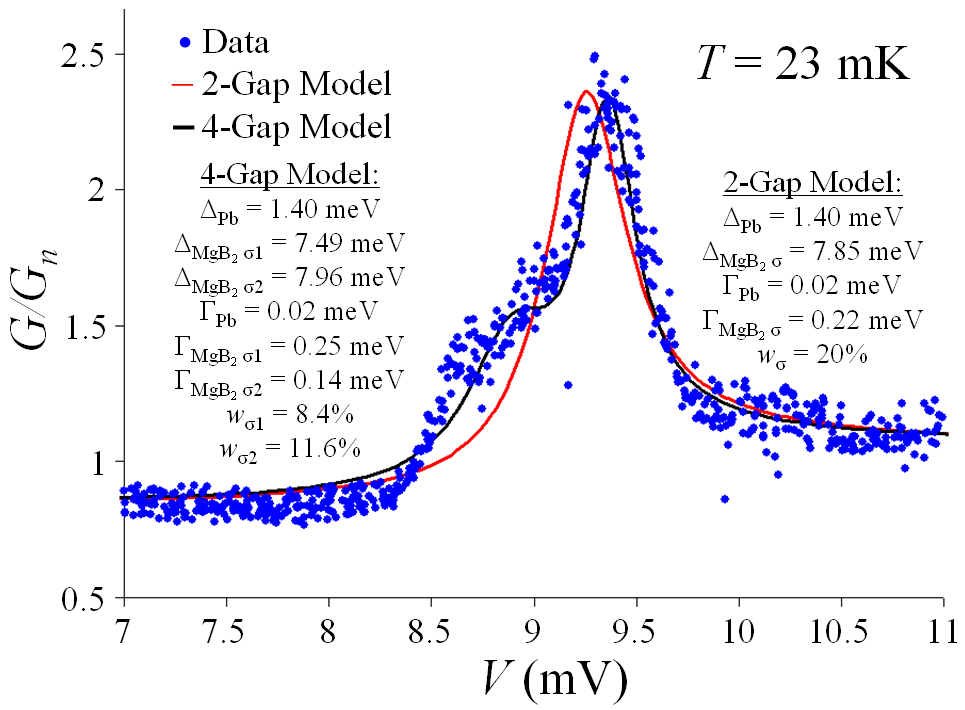}
\caption{ \label{Figure10}(color online) $\sigma$ gap for the MgB$_2$/I/Pb
``columnar'' junction. A series of features, including a prominent shoulder at
a lower voltage than the main peak, indicate that the $\sigma$ gap is more
complex than a single broadened peak.}

\end{figure}

\begin{figure}[htbp]
		\includegraphics[scale=1]{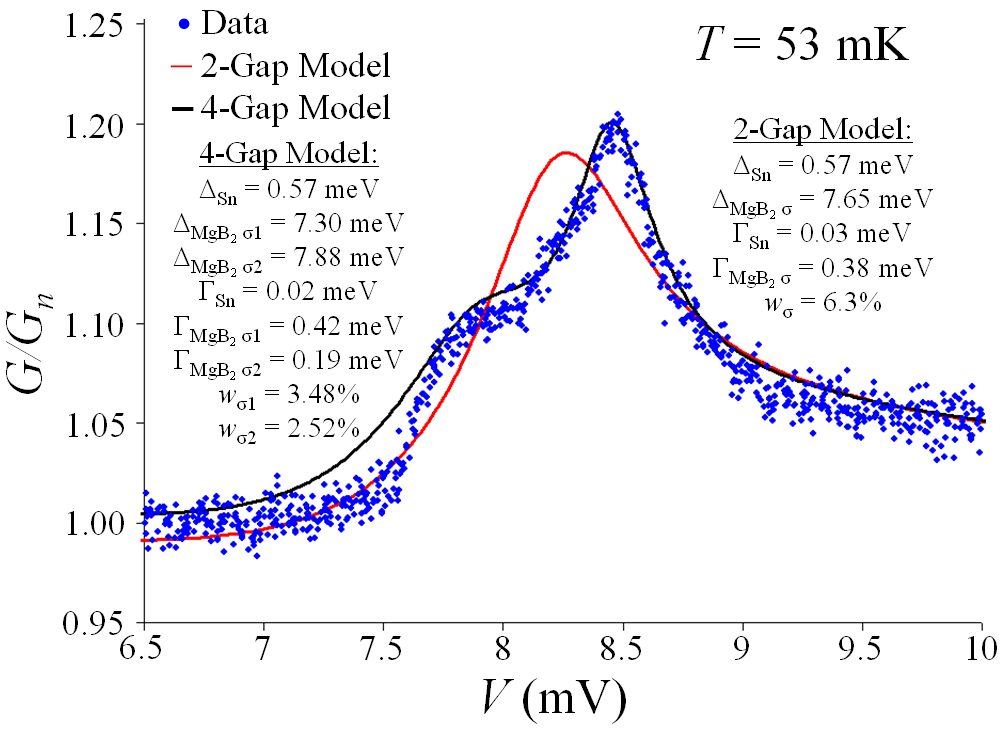}
\caption{ \label{Figure11}(color online) $\sigma$ gap for the MgB$_2$/I/Sn
``terraced'' junction. As above, a prominent shoulder appears at a voltage
below that of the main peak.}

\end{figure}

The ``terraced'' MgB$_2$/I/Sn junction also exhibits a sharper peak at higher
energies, and a prominent lower-energy shoulder. However, since $w_{\sigma}$ is
$\sim6\%$ (in contrast to the ``columnar'' junction's $\sim20\%$), the peaks
are less pronounced. As shown in Figure \ref{Figure11}, a single $\sigma$ gap
is far from adequate for representing this data, while two $\sigma$ sub-peaks
models the data reasonably well.

\begin{table*}
\caption{\label{ExptVsTheoryTable}Comparison of energy gap values of MgB$_2$
derived from fits to experimental data vs. peaks in theoretical density of
states calculations. Experimental uncertainties are estimated by the range of
gap values for which a 4-gap fit may produce reasonable agreement with the
data. Theoretical values are based on the center of each peak. Values for
$\Delta_{\sigma}$ for Theory (c) are approximate, as it is not easily separated
into two sub-peaks.}
\begin{ruledtabular}
\begin{tabular}{ccccc}
Feature & ``Columnar'' &  ``Terraced'' & Theory & Theory \\
 & Data & Data & (a) \cite{choi-2002-418} & (c) \cite{PhysRevB.87.024505} \\
\hline
$\Delta_{\pi\,1}$ (meV) & 1.78$\pm$0.02 & 1.77$\pm$0.04 & 1.51 & 1.3\\
$\Delta_{\pi\,2}$ (meV) & 2.32$\pm$0.1 & 2.3$\pm$0.2 & 2.06 & 2.2\\
$\Delta_{\sigma\,1}$ (meV) & 7.49$\pm$0.3 & 7.30$\pm$0.2 & 6.61 & $\sim$8.2\\
$\Delta_{\sigma\,2}$ (meV) & 7.96$\pm$0.05 & 7.88$\pm$0.05 & 7.13 & $\sim$8.7\\
\end{tabular}
\end{ruledtabular}
\end{table*}

\section{Discussion}

We have summarized our results in Table \ref{ExptVsTheoryTable}. Our data shows
that our experiments can resolve features down to less than 0.5meV apart. As
noted earlier, features in $NS$ conductance data are expected to be thermally
smeared by $\sim2k_{B}T$. This corresponds to 0.69meV at 4K, and 0.17meV at 1K.
$SIS'$ conductance data should be even sharper. This suggests our results are
not thermally limited.

Scattering will also limit the ability to resolve features within the energy
gaps. The scattering rate $\gamma$ may be calculated from $\gamma>\sqrt{\left
\langle \Delta \right \rangle \delta \Delta}$ \cite{PhysRevB.69.056501}, where
$\left \langle \Delta \right \rangle$ is the average order parameter, and
$\delta \Delta$ is the variation of the order parameter over the Fermi surface.
If we equate these with the average energy gap value and the resolution of our
energy gap data, respectively, we find a scattering rate on the order of 1 meV.
This implies a mean free path beyond 300 nm. Since this distance is on the same
order as irregularities in the film surface (\cite{ISI:000315667500054}), it is
surprising to observe this energy gap substructure, even with extremely clean
samples. Nevertheless, prior tunneling spectroscopy experiments at temperatures
from 7.0K to 1.8K have exhibited such resolved
features\cite{ncomms1626,ISI:000315667500054}, in addition to our experimental
results presented here.

Table \ref{ExptVsTheoryTable} also illustrates a significant discrepancy
between our experimental results and those from theory (a): our energy gap
values are consistently $\sim$10\% higher. There are several reasonable
explanations for this.

Nearly half of this difference may be accounted for by considering the effect
of strain (caused by different thermal expansion coefficients, as the sample is
cooled after growing the film) on energy gaps. It has been found experimentally
\cite{PhysRevLett.93.147006} that $T_c$ is approximately 41.5K for MgB$_2$
films on SiC, rather than the conventional value of 39.4K.

A key prediction of BCS theory is the value of the energy gap at zero
temperature \cite{nla.cat-vn1834745}:
\smallskip 
\begin{eqnarray}
\label{equation4}
E_g(0)=2\Delta(0)=3.528 k_B T_c
\end{eqnarray}

\noindent Therefore, elevating $T_c$ from 39.4K to 41.5K provides a roughly
5.3\% increase in the energy gap values. Similar experiments using similar
samples also found the sigma gap significantly elevated when using SiC as the
substrate for the MgB$_2$ film, rather than MgO (which has a much smaller
mismatch in expansion coefficients) \cite{ISI:000315667500054, ncomms1626}.

The remaining discrepancy may be due to using thin films for the experimental
realization of this measurement, rather than the periodic boundary conditions
used in calculations. Additional systematic differences between experiment and
theory are known to exist.

\section{Conclusion}

We have performed high-resolution tunneling measurements of low-transparency
MgB$_2$ tunnel junctions using ``terraced,'' ``columnar,'' and $c$-axis
geometries, at low (4K) to very low (23 mK) temperatures. With these
measurements, we have probed the substructures within the $\pi$ and $\sigma$
gaps of MgB$_2$.

Within the subgap, we observed very sharp peaks that identify, to high
precision, the values of the energy gaps of the junction counterelectrodes (Pb
and Sn). These lead us to conclude that the substructures seen in the $\pi$ and
$\sigma$ gaps are due to MgB$_2$, consistent with prior reported measurements
\cite{ncomms1626, ISI:000315667500054, chen:012502,10.1134_1.1780557, 5643942}.

Using a simplified two-band and four-band model with variable gap weights and
broadening factors, we demonstrate how these sub-structures illustrate the need
to go beyond a two gap model.

\begin{acknowledgments}

This research has been supported by a grant in aid from Sigma Xi, the
Scientific Research Society.

R.C.R. acknowledges partial support from National Science Foundation Grant \#
DMF-1206561 and Q.L. acknowledges support from DOE DE-FG02-08ER46531 (Q.L.).
The work at Temple University was supported by ONR under Grant No.
N00014-13-1-0052.
\end{acknowledgments}

\bibliography{Beyond2Gap_CarabelloEtAl_Refs}

\begin{thebibliography}{38}
\bibitem[{Not()}]{Note_NI9215}
\bibinfo{note}{For the MgB$_2$/I/Pb ``columnar'' junction, a NI 9215 DAQ was
  used. Its ports are not isolated; therefore, more averaging was necessary
  \textemdash though still reasonable considering the degree of oversampling.}
\end{thebibliography}


\end{document}